\documentclass[twocolumn,aps,showpacs,prl,superscriptaddress]{revtex4} 

\usepackage{sidecap}
\usepackage{ulem}
\usepackage{epsfig}
\usepackage{amsmath,amssymb,amsthm}
\usepackage{graphicx}
\usepackage{bm}
\usepackage{color}

\DeclareMathAlphabet{\mathpzc}{OT1}{pzc}{m}{it}

\begin{document}

\renewcommand{\textfraction}{0.00}


\newcommand{\vAi}{{\cal A}_{i_1\cdots i_n}} 
\newcommand{\vAim}{{\cal A}_{i_1\cdots i_{n-1}}} 
\newcommand{\vAbi}{\bar{\cal A}^{i_1\cdots i_n}}
\newcommand{\vAbim}{\bar{\cal A}^{i_1\cdots i_{n-1}}}
\newcommand{\htS}{\hat{S}} 
\newcommand{\htR}{\hat{R}}
\newcommand{\htB}{\hat{B}} 
\newcommand{\htD}{\hat{D}}
\newcommand{\htV}{\hat{V}} 
\newcommand{\cT}{{\cal T}} 
\newcommand{\cM}{{\cal M}} 
\newcommand{\cMs}{{\cal M}^*}
\newcommand{\vk}{\vec{\mathbf{k}}}
\newcommand{\bk}{\bm{k}}
\newcommand{\kt}{\bm{k}_\perp}
\newcommand{\kp}{k_\perp}
\newcommand{\km}{k_\mathrm{max}}
\newcommand{\vl}{\vec{\mathbf{l}}}
\newcommand{\bl}{\bm{l}}
\newcommand{\bK}{\bm{K}} 
\newcommand{\bb}{\bm{b}} 
\newcommand{\qm}{q_\mathrm{max}}
\newcommand{\vp}{\vec{\mathbf{p}}}
\newcommand{\bp}{\bm{p}} 
\newcommand{\vq}{\vec{\mathbf{q}}}
\newcommand{\bq}{\bm{q}} 
\newcommand{\qt}{\bm{q}_\perp}
\newcommand{\qp}{q_\perp}
\newcommand{\bQ}{\bm{Q}}
\newcommand{\vx}{\vec{\mathbf{x}}}
\newcommand{\bx}{\bm{x}}
\newcommand{\tr}{{{\rm Tr\,}}} 
\newcommand{\bc}{\textcolor{blue}}

\newcommand{\beq}{\begin{equation}}
\newcommand{\eeq}{\end{equation}} 
\newcommand{\ee}{\end{equation}}
\newcommand{\bea}{\begin{eqnarray}} 
\newcommand{\eea}{\end{eqnarray}}
\newcommand{\beqar}{\begin{eqnarray}} 
\newcommand{\eeqar}{\end{eqnarray}}
 
\newcommand{\half}{{\textstyle\frac{1}{2}}} 
\newcommand{\ben}{\begin{enumerate}} 
\newcommand{\een}{\end{enumerate}}
\newcommand{\bit}{\begin{itemize}} 
\newcommand{\eit}{\end{itemize}}
\newcommand{\ec}{\end{center}}
\newcommand{\bra}[1]{\langle {#1}|}
\newcommand{\ket}[1]{|{#1}\rangle}
\newcommand{\norm}[2]{\langle{#1}|{#2}\rangle}
\newcommand{\brac}[3]{\langle{#1}|{#2}|{#3}\rangle} 
\newcommand{\hilb}{{\cal H}} 
\newcommand{\pleft}{\stackrel{\leftarrow}{\partial}}
\newcommand{\pright}{\stackrel{\rightarrow}{\partial}}

\title{Radiative energy loss in a finite dynamical QCD medium}
\author{Magdalena Djordjevic}
\affiliation{Physics Department, The Ohio State University,
Columbus, OH 43210, USA}
\author{Ulrich Heinz}
\affiliation{Physics Department, The Ohio State University,
Columbus, OH 43210, USA}
\affiliation{CERN, Physics Department, Theory Division, CH-1211 Geneva 23, 
Switzerland}

\begin{abstract}
The radiative energy loss of a quark jet traversing a finite size QCD medium 
with dynamical constituents is calculated to first order in opacity. Although 
finite size corrections reduce the energy loss relative to an infinite 
dynamical QCD medium, under realistic conditions it remains significantly 
larger than in a static medium. Quantitative predictions of jet suppression 
in relativistic heavy ion collisions must therefore account for the dynamics 
of the medium's constituents. Finite size effects are shown to induce a 
non-linear path length dependence of the energy loss. Our results suggest a 
simple general mapping between energy loss expressions for static and 
dynamical QCD media.
\end{abstract}

\date{\today} 
 

\maketitle

{\it 1.} Studying the suppression of high transverse momentum 
hadrons is a powerful tool to map out the density of a QCD plasma created 
in ultrarelativistic heavy ion collisions \cite{Gyulassy_2002,%
Gyulassy:1990bh}. Since this suppression (called jet 
quenching) results from energy loss of fast partons moving through the 
plasma \cite{MVWZ:2004,BDMS,BSZ,KW:2004}, quantitative 
jet quenching predictions require reliable energy loss calculations.

In the majority of currently available studies the medium-induced 
radiative energy loss is computed by assuming that the QCD medium 
consists of randomly distributed {\it static} scattering centers 
(``static QCD medium''). We recently calculated \cite{DH_Inf}, at 
leading order in opacity, the heavy quark radiative energy loss in 
an infinite QCD medium consisting of {\it dynamical} constituents 
and found that the energy loss increases by almost a factor 2 relative 
to an equally dense static medium. However, this calculation was 
performed in the Bethe-Heitler limit which is well known \cite{DG_Ind} 
to overpredict radiative energy loss since it does not include coherence 
and finite size effects. As the medium created in heavy ion collisions 
has finite size, it is essential to explore how the qualitative conclusions 
obtained in \cite{DH_Inf} change once such effects are included. We find 
that finite size and coherence effects decrease the radiative energy loss 
per unit path length more strongly in a dynamical than in a static medium, 
reducing the energy loss ratio between equally dense dynamical 
and static media. Still, the ratio remains significantly larger than unity 
even if the medium is finite, showing that for quantitative predictions of
radiative energy loss it is important to account for the dynamic 
nature of the QCD medium's constituents. 

{\it 2.} We briefly outline the computation of the medium induced 
radiative energy loss for a heavy quark to first order in opacity. 
We consider a QCD medium of size $L$ and assume that the heavy quark 
is produced at time $x_0=0$ at the left edge of the medium, traveling 
right. Collisions with partons in the medium induce the radiation of 
gluons, causing the quark to lose energy. The radiative energy loss 
rate can be expanded in the number of scattering events suffered by the 
heavy quark. This is equivalent to an expansion in powers of the opacity. 
For a finite medium, the opacity is given by the product of the medium
density with the transport cross section, integrated along the path of 
the heavy quark. The leading (first order) contribution corresponds to 
one collisional interaction with the medium, accompanied by the emission 
of a single gluon. This is the process we compute.

To introduce the finite size of the medium we start from the approach 
described in \cite{Zakharov} and follow the procedure used in 
\cite{MD_Coll}. The medium extends for a length $L$ from the production
point of the energetic heavy quark, and the collisional interaction
inducing the radiation of a gluon occurs after a distance $l<L$ inside 
the medium.     

\begin{SCfigure*}
\epsfig{file=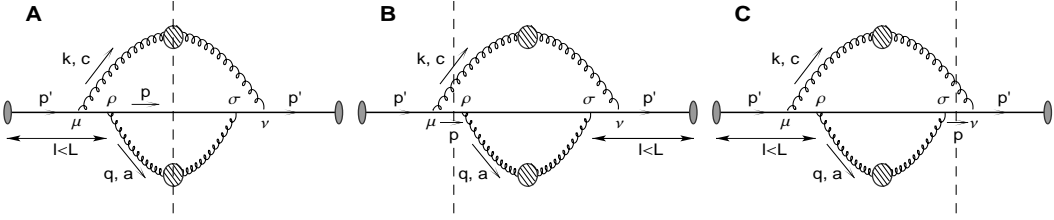,width=5.5in,height=1.2in,clip=5,angle=0}  
\caption{Three typical Feynman dia\-grams contributing to the quark radiative 
energy loss in a finite size dynamical QCD medium at first order in opacity.
See text for discussion.}
\label{Diag_Rad}
\end{SCfigure*}

As in \cite{DH_Inf}, we describe the medium by a thermalized quark-gluon 
plasma at temperature $T$ and zero baryon density, with $n_f$ effective 
massless quark flavors in equilibrium with the gluons. Three typical 
Feynman diagrams contributing to the radiative quark energy loss at first 
order in opacity are shown in Fig.~\ref{Diag_Rad}. The diagrams are evaluated 
in finite temperature field theory \cite{Kapusta,Le_Bellac}, using Hard 
Thermal Loop (HTL) resumed propagators \cite{Le_Bellac} for all gluons. A 
full account of the calculation will be presented elsewhere \cite{FinCalc}. 
A flavor of what it involves is given by Fig.~\ref{Diag_Rad}. The elliptic 
blob represents a source $J$ which at time $x_0$ produces an energetic quark 
with momentum $p'$. In diagram \ref{Diag_Rad}A the produced quark is on-shell. 
It first radiates a gluon with momentum $k=(\omega,k_z,\bk)$ and then 
exchanges a virtual gluon of momentum $q=(q_0,q_z,\bq)$ with a parton in the 
medium, finally emerging (at the dashed line denoting an on-shell cut through 
the amplitude represented by the diagram) with (measured) momentum 
$p=(E,p_z,\bm{p})$ \cite{fn1}. Since the energetic quark produced by the 
source $J$ can be off-shell, we also have contributions such as those in 
Figs.~\ref{Diag_Rad}B and \ref{Diag_Rad}C. Amplitudes \ref{Diag_Rad}B 
and \ref{Diag_Rad}C interfere with amplitude \ref{Diag_Rad}A, leading to the 
appearance of LPM-like effects once all relevant contributions are summed. 
The present calculation differs from that in \cite{DG_Ind} by the use of HTL 
gluon propagators to describe the interaction of the quark with the medium, 
and from that in Ref.~\cite{DH_Inf} by allowing the jet to be on- or off-shell 
and restricting the vertices corresponding to gluon exchange to be located 
inside the medium, i.e. at $l<L$. We use the same kinematic approximations 
as in \cite{DG_Ind,DH_Inf}; accordingly, the gluon propagators for 
exchanged gluons in Fig.~\ref{Diag_Rad} contribute only for space-like momenta 
($q_0 < |\vq|$) and those for radiated gluons only for time-like momenta 
($\omega> |\vk|$) \cite{MD_Coll,DH_Inf}. We also assume that $J$ changes 
slowly, i.e.  $J(p') \approx  J(p)$~\cite{GLV}.

Explicit calculation of all 24 diagrams contributing to first order in 
the opacity \cite{FinCalc} yields the following expression for the 
fractional radiative energy loss ($\mu=gT\sqrt{N_c/3+N_f/6}$ is the
Debye screening mass and parametrizes the density of the medium):
\beqar
\label{DeltaEDyn}
\frac{\Delta E_{\mathrm{dyn}}}{E} 
&=& \frac{C_R \alpha_s}{\pi}\,\frac{L}{\lambda_\mathrm{dyn}}  
    \int dx \,\frac{d^2k}{\pi} \,\frac{d^2q}{\pi} \, 
    \frac{\mu^2}{\bq^2 (\bq^2{+}\mu^2)}\,
\\   
&& \hspace*{-0.8in}\left(1-\frac{\sin({\frac{(\bk{+}\bq)^2+\chi}{x E^+} \, L})} 
    {\frac{(\bk{+}\bq)^2+\chi}{x E^+}\, L}\right) 
    \frac{2(\bk{+}\bq)}{(\bk{+}\bq)^2{+}\chi}
    \left(\frac{(\bk{+}\bq)}{(\bk{+}\bq)^2{+}\chi}
    - \frac{\bk}{\bk^2{+}\chi}
    \right) \!.
\nonumber
\eeqar
Here $\lambda_\mathrm{dyn}^{-1} \equiv C_2(G) \alpha_s T = 3 \alpha_s T$ 
defines the ``dynamical mean free path'' \cite{DH_Inf}, $\alpha_s = 
\frac{g^2}{4 \pi}$ is the strong coupling constant, 
and $C_R{=}\frac{4}{3}$. Further, 
$\chi\equiv M^2 x^2 + m_g^2$ where $x$ is the longitudinal momentum 
fraction of the heavy quark carried away by the emitted gluon and 
$m_g=\frac{\mu}{\sqrt{2}}$ is the effective mass for gluons with 
hard momenta $k\gtrsim T$.

Similar to the infinite medium studied in \cite{DH_Inf}, each individual 
diagram contributing to the energy loss in a finite dynamical medium 
diverges logarithmically in the limit of zero transverse momentum exchange, 
$\bq{\,\to\,}0$ \cite{FinCalc}. In a dynamical QCD medium both transverse 
and longitudinal gluon exchange contribute to radiative energy 
loss \cite{Wang_Dyn}; while Debye screening renders the longitudinal gluon 
exchange infrared finite, transverse gluon exchange causes a logarithmic 
singularity due to the absence of magnetic screening \cite{Le_Bellac}. 
Remarkably, this singularity is found to cancel in the sum over all diagrams 
\cite{FinCalc}, naturally regulating the energy loss rate.

{\it 3.} We can compare the radiative energy loss rate in a dynamical medium 
(\ref{DeltaEDyn}) to the analogous result for a static medium. One can rewrite 
the DGLV expression \cite{DG_Ind} for the first order radiative energy loss in 
a static QGP, $\Delta E_{\mathrm{stat}}/E$, in the same form as 
Eq.~(\ref{DeltaEDyn}), except for two simple substitutions: 
(1) $\lambda_\mathrm{dyn}$ is replaced by the ``static mean free path'' 
$\lambda_\mathrm{stat}$, defined by \cite{WHDG,DH_Inf}
\beqar
\label{lambda_stat}
\frac{1}{\lambda_\mathrm{stat}}=\frac{1}{\lambda_{g}}+\frac{1}{\lambda_{q}}=
6 \frac{1.202}{\pi^2} \frac{1{+}\frac{n_f}{4}}{1{+}\frac{n_f}{6}} 3 \alpha_s T
= c(n_f) \frac{1}{\lambda_\mathrm{dyn}} ,
\eeqar
where $c(n_f) \equiv 6 \frac{1.202}{\pi^2} \frac{1+n_f/4}{1+n_f/6}$
is a slowly increasing function of $n_f$ that varies between 
$c(0)\approx 0.73$ and $c(\infty)\approx1.09$. For $n_f=2.5$ (see below)
$c(2.5) \approx 0.84$. (2) The effective cross section under the
integral (\ref{DeltaEDyn}) for the energy loss rate is replaced as
\beq
\label{diff_cross}
  \left[ \frac{\mu^2}{\bq^2 (\bq^2{+}\mu^2)} \right]_\mathrm{dyn}
  \mapsto
   \left[\frac{\mu^2}{(\bq^2{+}\mu^2)^2}\right]_\mathrm{stat}.
\eeq
Taken together, these differences will be seen to cause a significant increase 
of the heavy quark energy loss rate in dynamical compared to static QCD media. 

These two simple replacements are identical to those found in the Bethe-Heitler 
limit \cite{DH_Inf}. The simplicity of this substitution rule is surprising, 
given the complexity of the calculations and their different structure for 
static \cite{DG_Ind} and dynamical \cite{FinCalc} media. (Remember the infrared 
divergences in the dynamical case which cancel only after summing all 24 
diagrams but don't arise at all in the static case.) The integrands in 
Eq.~(\ref{DeltaEDyn}) and its static analog are significantly different from 
the corresponding ones in the Bethe-Heitler limit \cite{DG_Ind}, giving 
rise (as we will see) to a different energy dependence of the dynamic/static 
energy loss ratio. Nonetheless the same simple substitution rule is found to 
apply, suggesting a possibly general mapping between static and dynamic QCD 
media. 

%

\begin{SCfigure*}
\epsfig{file=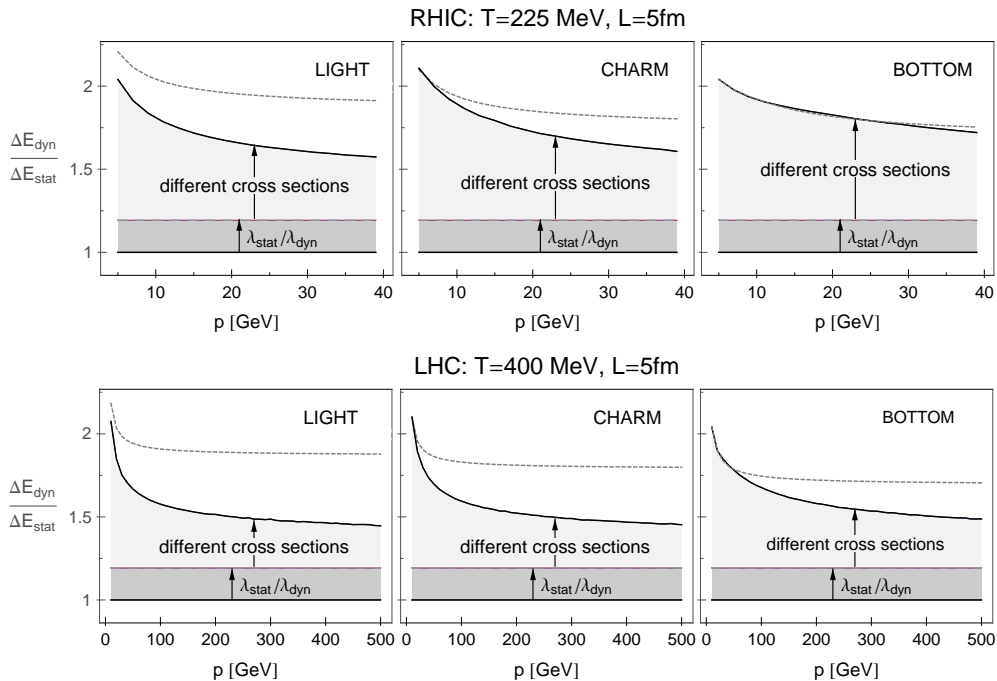,width=5.3in,clip=5,angle=0}
\caption{Ratio of the radiative energy loss in finite dynamical and 
static QCD media of length $L=5$\,fm for light, charm and bottom quarks 
(left, center, and right panels, respectively), as a function of initial 
quark momentum $p$. Top row: RHIC conditions (average medium temperature 
$T=225$\,MeV). Bottom row: LHC conditions ($T=400$\,MeV). The dashed 
curves show the corresponding energy loss ratio in an infinite QCD medium 
for comparison.
}
\label{ratioRHIC_LHC}
\end{SCfigure*}
%

The study presented here considers a finite, optically thin dynamical QCD 
medium (QGP), extending the DGLV approach \cite{DG_Ind} to include parton 
recoil. In this sense it is complementary to the work by Arnold, Moore and 
Yaffe \cite{AMY} who study energy loss in an infinite, optically thick QGP. 
We note that the AMY approach \cite{AMY} yields the same form (\ref{diff_cross}) 
for the effective cross section in a dynamical QCD medium as found here (see 
also~\cite{Aurenche}), supporting our conjecture above.

{\it 4.} We now highlight finite size effects in a dynamical QCD medium, 
to first order in opacity, with a few numerical results for 
radiative energy loss. In Fig.~\ref{ratioRHIC_LHC} we show the ratio of 
the radiative energy loss rates in equally dense dynamical and static QCD 
media as a function of the initial energy of the fast quark, under 
RHIC and LHC conditions. In both cases a medium of length $L=5$\,fm, a 
constant value of $\alpha_s{\,=\,}0.3$, and a chemically equilibrated QGP 
with $n_f{\,=\,}2.5$ effective light quark flavors is assumed. The
light quark mass is assumed to be dominated by the thermal mass, 
$M_q{\,=\,}\mu/\sqrt{6}$, where $\mu$ is the Debye screening mass. For the 
charm and bottom masses we use $M_c{\,=\,}1.2$\,GeV and 
$M_b{\,=\,}4.75$\,GeV, respectively. For Au+Au collisions at top RHIC 
energies we assume an average medium temperature of $T=225$\,MeV, for Pb+Pb 
at the LHC we take $T=400$\,MeV. 

In all cases, the energy loss is seen to be significantly larger in the 
dynamical than in the static medium. A common factor to all situations 
(large and small jet quark masses, hotter and cooler media, finite and 
infinite media) is the $\mathcal{O}(20\%)$ increase of the energy loss in 
dynamical media arising from the shorter mean free path 
$\lambda_\mathrm{dyn}\approx0.84\lambda_\mathrm{stat}$. The additional 
increase arising from the change (\ref{diff_cross}) in cross section  
is larger; in the energy range shown in Fig.~\ref{ratioRHIC_LHC} it 
ranges from about 25\% to over 100\%, depending on medium temperature 
and the mass and energy of the fast quark. The reduction of the energy
loss ratio $\frac{\Delta E_\mathrm{dyn}}{\Delta E_\mathrm{stat}}$ by finite 
size corrections is seen to be larger for lighter quarks and larger jet 
energies. The smallest finite size corrections and, in the end, the biggest 
dynamical increase are seen for bottom quarks at RHIC. 

Furthermore, in \cite{DH_Inf} we found that dynamical medium effects are 
{\it largest} for light quarks and {\it decrease} with quark mass. Here we see 
the opposite ten\-den\-cy: after finite size correction the energy loss ratio 
$\frac{\Delta E_\mathrm{dyn}} {\Delta E_\mathrm{stat}}$ becomes {\it smallest} 
for light quarks, {\it increasing} with quark mass. Such behavior is 
important, since it may contribute toward understanding the observed large 
suppression of non-photonic electrons in central Au+Au collisions at 
RHIC \cite{e_suppression}.

\begin{figure*}
\includegraphics[bb=100 25 830 240,width=\linewidth,clip=]{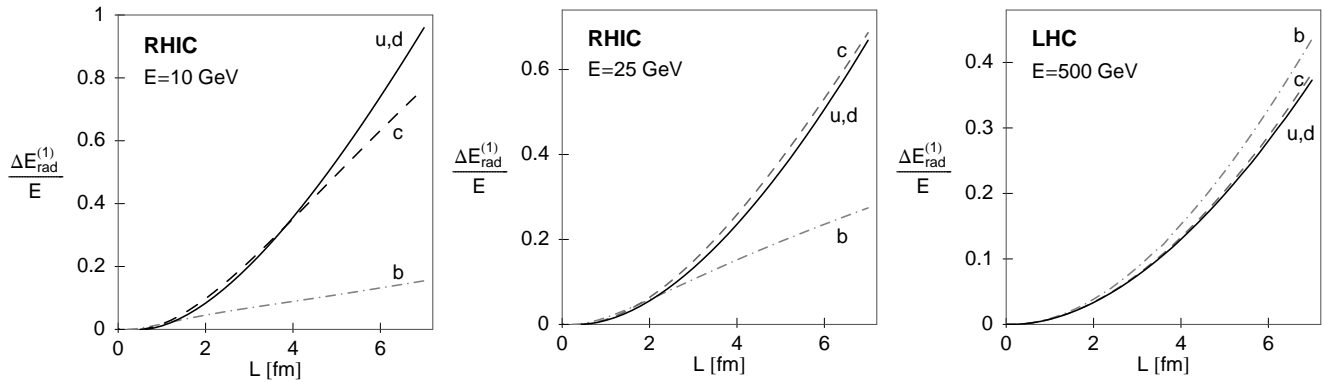}
\caption{First order fractional radiative energy loss as a function of 
medium thickness $L$ for initial jet energies $E=10,\,25$, and
500\,GeV  (left, center, and right panels, respectively). The two left 
panels correspond to RHIC conditions, the right panel to LHC conditions. 
Solid, dashed and dot-dashed lines describe light, charm, and bottom quark
energy loss, respectively. }
\label{Ldependence}
\end{figure*}

Figure~\ref{Ldependence} shows that the strength of the finite size 
corrections correlates with the dependence of the fractional energy loss 
on the thickness $L$ of the medium. In the Bethe-Heitler limit studied for 
infinite media in Ref.~\cite{DH_Inf}, quarks of all masses and energies 
lose energy at fixed rate $\frac{\Delta E}{\Delta z}$, resulting in a linear 
dependence of the radiative energy loss on the length $L$ traveled by the 
quark. In contrast, Fig.~\ref{Ldependence} shows a non-linear $L$-dependence
that becomes perfectly quadratic (corresponding to the deep 
Landau-Pomeranchuk-Migdal (LPM) limit \cite{LPM}) for large jet energy 
(see Eq.~(\ref{E_assymp}) below). The weakest deviations from the linear 
Bethe-Heitler $L$-dependence are seen for low-energy bottom quarks at RHIC, 
where Fig.~\ref{ratioRHIC_LHC} (top row, right panel) also shows 
the smallest finite size correction. The $L$-dependence is closest to quadratic 
for light quarks and for very energetic charm and bottom quarks at the LHC 
where also the finite size effects are largest. This shows that the finite 
size corrections implemented in the present calculation simulate the 
destructive effects of LPM interference in an infinite medium \cite{BDMS}. 
This behavior is expected \cite{GLV} since the nuclear medium has finite 
dimensions that may be small compared to the jet radiation coherence length, 
especially in the case of light partons or high jet energies. Due to this, in 
finite size media the basic formation time physics developed by LPM \cite{LPM} 
leads to strong destructive interference effects on the quark quenching, 
as observed in Fig~\ref{Ldependence}. 

{\it 5.} We finally point out that, contrary to the first order study in the 
Bethe-Heitler limit \cite{DH_Inf} where the energy loss ratio 
%
%
$\frac{\Delta E_\mathrm{dyn}}{\Delta E_\mathrm{stat}}$ saturates for 
sufficiently large quark energies, this ratio keeps decreasing with 
increasing quark energy once finite size corrections are accounted for. 
In fact, one finds analytically \cite{FinCalc} that for asymptotically 
large jet energies Eq.~(\ref{DeltaEDyn}) reduces to
\beqar
\label{E_assymp}
  \frac{\Delta E_{\mathrm{dyn}}}{E} &\approx& \frac{C_R \alpha_s} {4} 
  \frac{L^2 \mu^2} {\lambda_\mathrm{dyn}} \ln\frac{4 E T}{\mu^2}\,, 
\eeqar
and that the energy loss ratio approaches
\beqar
\label{ratio_assymp}
  \lim_{E \rightarrow \infty} \frac{\Delta E_{\mathrm{dyn}}}
  {\Delta E_{\mathrm{stat}}}= \lim_{E \rightarrow \infty}
  \frac{\lambda_\mathrm{stat}} {\lambda_\mathrm{dyn}} 
  \frac{\ln\frac{4 E T}{\mu^2}}{\ln\frac{4 E T}{\mu^2}{-}1} 
  = \frac{\lambda_\mathrm{stat}} {\lambda_\mathrm{dyn}} \,.
\eeqar
The static approximation thus becomes valid for asymptotically large 
jet energies.  

{\it 6.} In summary, we have presented a calculation to first order in 
opacity of the radiative energy loss of a fast quark traveling through 
a finite dynamical QCD medium. Finite size effects are found to be most 
important in the ultrarelativistic limit and they effectively reproduce the 
effects of destructive Landau-Pomeranchuk-Migdal interference. The 
calculation suggests the possibility of a general mapping between the 
energy loss expressions for static and dynamical media, which we 
conjecture to carry over to higher-order calculations. It also shows that 
the approximation of the medium by a random distribution of static scatterers 
becomes valid in the limit of asymptotically large jet energies once finite 
size and LPM interference effects are taken into account. For realistic jet 
energies and medium temperatures reachable at RHIC and LHC, however, parton 
recoil in the medium must be accounted for and leads to a large (40-70\%) 
increase of radiative energy loss when compared with an equally dense static 
medium. This effect is largest for bottom quarks at RHIC which may be important
for understanding the observed large suppression of non-photonic electrons
in central Au+Au collisions at RHIC \cite{e_suppression}.   

{\it Acknowledgments:}
Valuable discussions with Eric Braaten, 
Miklos Gyulassy, and Yuri Kovchegov are gratefully acknowledged. This work 
was supported by the U.S. Department of Energy, grant DE-FG02-01ER41190.



\begin{thebibliography}{80}
\bibitem{fn1}
  We use the same notation as in \cite{DH_Inf}.
\end{thebibliography}
\end{document}